*Gold nanoparticle-based brachytherapy enhancement in choroidal melanoma using a full Monte Carlo model of the human eye*

*Somayeh Asadi[1*], Mehdi Vaez-zadeh[1], S. Farhad Masoudi,[1] Faezeh Rahmani[1], Courtney Knaup[2], and Ali S. Meigooni[2]*

[1]*Department of Physics, K.N. Toosi University of Technology, Tehran, Iran*
[2]*Comprehensive Cancer Centers of Nevada, Las Vegas, Nevada, USA*

The effects of gold nanoparticles in [125]I brachytherapy dose enhancement on choroidal Melanoma are examined using the Monte Carlo simulation technique. Usually, Monte Carlo ophthalmic brachytherapy dosimetry is performed in a water phantom. However, here, the compositions of human eye have been considered instead of water. Both human eye and water phantoms have been simulated with MCNP5 code. These simulations were performed for a fully-loaded 16 mm COMS eye plaque containing 13 [125]I seeds. The dose delivered to the tumor and healthy tissues have been calculated in both phantoms, with and without GNPs. The results indicates that the dose to the tumor in an eye-ball implanted with COMS plaque increases with increasing GNPs concentration inside the target. Therefore, the required irradiation time for the tumors in the eye is decreased by adding the GNPs prior to treatment. As a result, the dose to healthy tissues decreases when the irradiation time is reduced. Furthermore, a comparison between the simulated data in an eye phantom made of water and eye phantom made of human-eye composition, in the presence of GNPs shows the significance of utilizing the composition of eye in ophthalmic brachytherapy dosimetry. Normally, the radiation therapy of cancer patients is designed to deliver a required dose to the tumor while sparing the surrounding healthy tissues. The results demonstrated that the use of GNPs enable us to overcome this challenge. Also, defining the eye composition instead of water will leads to more accurate calculations of GNPs radiation effects in ophthalmic brachytherapy dosimetry.

[*]Corresponding Author: Somayeh Asadi, *E-mail address*: s_asadi@sina.kntu.ac.ir

## I. Introduction

Ocular melanoma or more specifically, uveal melanoma is a malignant tumor which can arise from the melanin-producing cells or melanocytes residing within the uvea. These kinds of tumors have the highest rate of metastasis of any intraocular cancer. The method of treatment is determined according to the type of the cancer and the rate of its progress.[1-3] Enucleation, local resection and radiation therapy are the most common methods of treatment for ocular melanoma[4]. In radiation therapy, penetrating radiation like X-ray, gamma-ray, alpha and beta beams are used. These radiation types could be emitted from either a radiation apparatus such as linear accelerator or sealed radioactive sources (also known as brachytherapy sources) or from radio-labeled substances. Brachytherapy is a form of radiation therapy which involves placing small sealed radioactive seeds inside or adjacent to the tumor.[5-6] Studies have shown that for most eye melanomas, the methods of using removable ophthalmic plaques loaded with brachytherapy sources are as effective as surgery (Enucleation).[7-9]

Several isotopes, including $^{125}$I and $^{103}$Pd are used in the brachytherapy of choroidal Melanoma.[10-12] Increasing the dose to the tumor, while decreasing the dose to the normal tissues remains as one of the biggest challenges in radiation therapy. Therefore, the design of the ophthalmic plaques and selection of the radionuclide for intraocular cancers has been the primary goal of the clinical studies for many investigators.[13-15] Several investigations based on Monte Carlo simulations or experimental techniques have been performed for dosimetry of choroidal Melanoma, using different ophthalmic plaques containing various brachytherapy sources[7-10, 13].

Presently, there is increased interest in the potential use of gold nanoparticles (GNPs) as a dose enhancer in resolving the noted challenges in the treatment cancer patients through ionization radiation.[15-16] In the context of such a treatment modalities, GNPs are considered to have advantageous characteristics such as biocompatibility, inertness and no reported obvious toxicity[17-18]. High photon interaction cross section of gold resulting from its high atomic number and electron density increases the possibility of dose absorption by GNPs. Because of the strong photoelectric absorption and secondary electron released by low energy gamma-ray or X-ray irradiation, GNPs can increase in dose enhancement factor (DEF) which will accelerate DNA strand breaks.[19-21] DEF is defined as the ratio of the absorbed dose by the tumor containing GNPs, to the absorbed dose by the tumor without these nanoparticles. In an in-vivo study in mice bearing subcutaneous EMT-6 mammary carcinoma, Hainfeld *et al.*[22] demonstrated that the presence of GNPs in the tumor will cause more absorbed dose by the cancerous cells than that of the healthy tissues. Irradiation stability and cytotoxicity of GNPs in human K562 cells have been investigated by Xiao-Dong Zhang *et al*[16] and the results indicated that GNPs do not deteriorate under high energy ray irradiation and showed concentration-dependent cytotoxicity. Also, the toxicity of the nanoparticles have been examined via different ways of assessing the cell viability like methyl thiazol tetrazolium (MTT) and Cell Titre-Glo™ luminescent cell viability assay.[16, 23] In some of the In-Vivo and In-Vitro studies[23-27], after intravenous injection of nanoparticles, transmission electron microscopy (TEM) demonstrated that GNPs are accumulated in clusters within the membrane bound vesicles and lysosomes. The tumor vasculature shows more transmission than the normal blood vessels and there is no lymphatic drainage in the tumor. So, due to the poorly formed tumor vasculature, the accumulation of unlabeled nanoparticles within the tumor can occur under passive targeting by increasing the effect of permeability and retention (EPR) [28-29]. Several other studies demonstrated the effects of GNPs in radiation treatment of the different tumors.[23, 30-38]

In the Monte Carlo study by Lechman *et al*[35], it is reported that the energy deposited by photoelectrons in the tumor area, is more important than the size and concentration of gold nanoparticles. However, in an independent project by Leung et al [36] using the MC technique, it was shows that gold nanoparticles with larger diameter and concentration increases the DEF in the tumor area. This means that to attain more precise results and knowing the parameters effective in cancer treatment by gold nanoparticles, more studies and examinations should be carried out.

Considerable advances have been made in the application of nanotechnology-based cancer therapy and numerous studies have been carried out in this field through Monte Carlo simulation and experimental studies. However, only limited studies have been reported that relate eye tumors. From these resources, Sheng Zhang *et al*[39] has studied the particles accumulation in the uveal tissue. In this report, the combination of nanoparticles of a suitable size with the ligands specified for the uveal melanoma cells has been found to be a good way for transferring these particles to the tumor area which would help them to stay in the melanoma tissue for a long time. In addition, they have reported that nanoparticles can escape through the uveal into the melanoma tissue with much higher accumulation than the micro-particles with optimal sizes of 100 nm to 300 nm. In a different study, Shin J. Kang *et al*[40] have evaluated the efficacy of subconjunctival nanoparticles carboplatin in the Murine Retinoblastoma treatment. However, to our knowledge, the impact of the nanoparticles on the radiation dosimetry of the eye has not yet been evaluated. Therefore, this project has been designed to evaluate the enhancement of the absorbed dose by the presence of the GNPs within the human eye considering both water and human eye compositions.

In this study we have investigated the application of GNPs in brachytherapy on the eye tumor by Monte Carlo simulations. These simulations are performed with the eye model phantom and water phantom (eye model filled with water ) to determine the dose enhancement factor (DEF). In addition, the significance of the eye model on the DEF calculation has been evaluated by comparing the dosimetry calculations in the presence of GNPs in both eye model phantom and water phantom. These simulations are performed using MCNP5 Monte Carlo code. Dosimetric characteristics of a single source were utilized to validate the accuracy of the Monte Carlo simulation technique. Also the dose distributions of the multi-source implant geometry in both eye and water phantoms with 16mm COMS standard eye-plaque loaded with $^{125}$I, model 6711 source (manufactured by GE Healthcare/Oncura) have been simulated.

## II. Material and Method
### A. Monte Carlo Simulation

In this project, the MCNP5 code[41] has been utilized to evaluate the effect of GNPs on dose enhancement for treatment of ocular melanoma with $^{125}$I eye-plaque therapy. This code utilizes a three-dimensional heterogeneous geometry and transports for photons and electrons in the energy range from 1 KeV to 1 GeV. For the simulations of this project, the default MCNP5 photon cross-section library, which is based upon release 8 of the ENDF/B-VI[42] has been utilized. The simulations were performed with the *F8 tally for the phantom dosimetry and *F6 tally for the air kerma simulations. The calculated data in MeV per incident particle for each section of the phantom was then converted to absorbed dose (i.e. Gy) or absorbed dose rate per air kerma strength of the source (i.e. cGy h$^{-1}$U$^{-1}$) by introducing appropriate conversion factors following the published guidlines.[43-44] Number of $7*10^7$ and $2*10^9$ histories have been followed to simulate the air kerma and phantom dosimetry, respectively, in order to achieve a relative statistical error of less than 1%. Table 1 shows the error propagation of the Monte Carlo simulation with MCNP5 in this project.

The effects of the nanoparticles were calculated by performing the simulation in the eye-model filled with the water and human eye composition, with and without the presence of the GNPs. Moreover, the geometric design of a 16mm COMS standard eye-plaque, loaded with 13

[125]I, model 6711 source (manufactured by GE Healthcare/Oncura) have been introduced in the simulations. The water phantom and the human eye globe geometries (Figure 1) were designed in a manner similar to the one described in our previous work.[45]

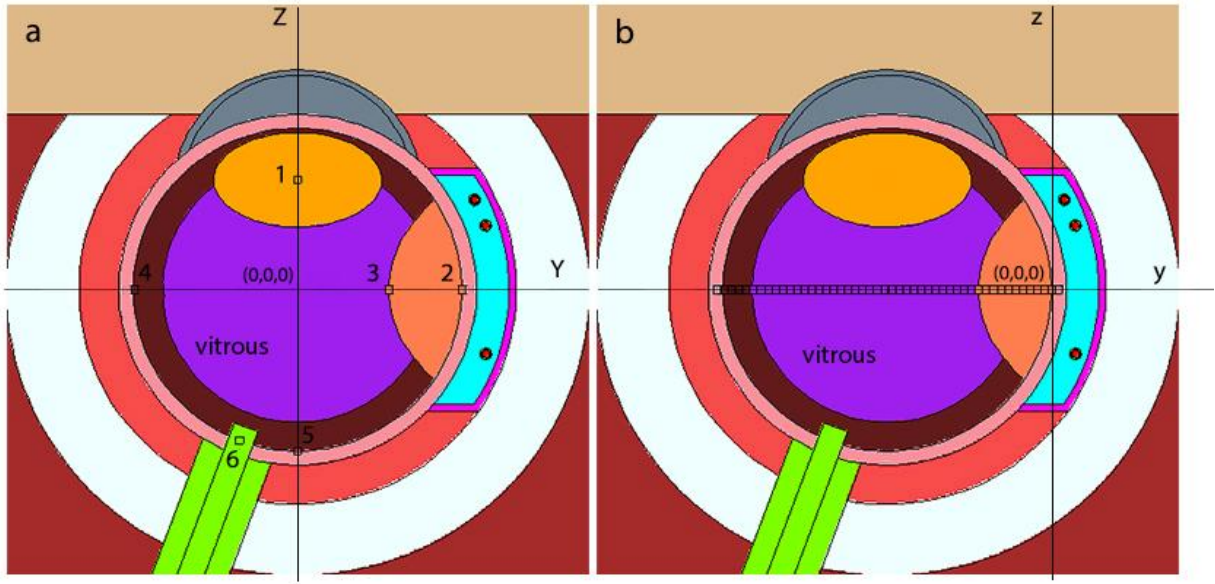

Figure 1: The longitudinal cross sectional diagram of the simulated human eye. In the left panel (a), the origin of the eye coordinate system is located at the center of the eye phantom. The voxels (numbered 1 to 6) indicate the lens, sclera, Tumor apex, opposite side, Macula and optic nerve, respectively. In right panel (b), the origin of the plaque coordinate is at the interior surface of the sclera. The tally cells on this figure are shown along the central axis of the plaque, starting from sclera near the plaque to the sclera at the opposite side of the eye.

As has been emphasized in the previous report, the geometry and characteristic of this simulated eye globe have been determined in a manner that the dimension and specification of main parts of the human eye could be in conformity with the medical data[44-53]. The chemical compositions and densities of different components of the eyeball tissues which were used in these simulations are given in Table2. Due to the nervous complexity of the Retina and the vascular complexity of the choroid, their density and chemical compositions were assumed to be the same as water. A brief description of the components and shapes of the eyeball used in these simulations are given in the following section.

The dimensions of the adult human eyeball are relatively constant and may vary by only one or two millimeters, from one person to another. Normally, an adult eyeball has an anterior to posterior diameter of 24 millimeters, but the vertical and horizontal diameters are approximately 23mm and 23.5 mm, respectively. Therefore, a spherical shell of 24.6 mm has been considered as an eye globe, during these simulations. In these simulations, the common volume between four concentric spheres with radii of 0.93, 1.03, 1.13 and 1.23 cm (centered at the center of the eye) are the introducer of Retina, Choroid and Sclera, (three primary layers of the eye), respectively. The Vitreous humor fills the space between the Lens and the smallest sphere (the inner surface of the Retina). The lens was defined as an ellipsoidal shape with equatorial diameters of 8mm and 9mm and a polar diameter of 0.25 cm. Also, the optic nerve has been defined as the volume between two concentric cylinders with diameters of 7 mm and 8 mm, which was located at the outer layer of the Sclera. The skull bone has been simulated by considering the volume between two concentric spheres (centered at the origin of the eye coordinate) with radii of 1.505 cm and 2.05 cm. In both water and human eye globe phantom, the array of 0.5×0.5×0.5 mm$^3$ voxels have been utilized for scoring the average energy deposition to different points in the eyeball (Figure1-B). We have examined that the use of smaller voxel size (i.e. 0.1×0.1×0.1 mm$^3$) will lead to similar results but with much longer run time on computer, in order to achieve similar statistical uncertainties. Therefore, this voxel size (i.e. 0.5×0.5×0.5 mm$^3$) utilized for the final data collection. These tally cells are selected along the central axis of the plaque, starting from sclera near the plaque to the sclera at the opposite side of the plaque.

A single source dosimetry of Model 6711 $^{125}$I source was used to validate the accuracy of the source and phantom geometry used in these Monte Carlo simulations. TG-43[54-55] recommended dosimetric characteristics of the single source (i.e. dose rate constant, radial dose function, and 2D anisotropy function) were compared with the published data for this validation process. These simulations were performed using a single source located at the center of a 30×30×30 cm$^3$ water phantom. To calculate the dose fall-off of the source, along its transverse axis with the smaller statistical fluctuations, toroid tally cells (torus-shaped cells) were selected. The major radii of these cells were chose to be in the range of 0.05 cm to 10 cm. The minor radii of the toroid cells, "R", varied as a functional of radial distance as; [R=0.008 cm for 0.05 < r ≤ 0.1; R=0.01 cm for 0.1 < r ≤ 1 cm; R=0.05 cm for 1 < r ≤ 5; and R=0.1 cm for 5 < r ≤ 10 cm].

In the Monte Carlo calculations of this project, the simulations for air kerma rates are performed with the tally cells for scoring the collisional kerma at various distances relative to the source center. The calculated air kerma rates have been scored in the toroid cells filled with the dry air and located at distances ranging from 0.05 cm to 20 cm in vacuum and F6 tally was used. The air kerma strength is defined as the product of the air kerma rate at a given distance from the source center in vacuum by the square of the distance, as defined in the TG-43U1 recommendation. The $S_K$ is independent from the distance. The air kerma strength was then calculated for all the distances and has been found to be constant from 1cm to 10 cm with relative statistical uncertainties less than 0.1%, so the average value in this region has been taken as the air kerma strength per history.

### B. Choroidal Melanoma

Choroidal Melanoma is one of the three kinds of the Uvea melanoma (choroidal, ciliary body and iris melanoma). Treatment of the intraocular tumor depends on its basal diameter and apical height. Plaque brachytherapy can be used for the treatment of tumors with the apical height of 2.5 to 10 mm and the basal diameter of 16 mm or less.

In this work, a choroidal melanoma tumor with the apical height of 5 mm (i.e. 6mm from the exterior surface of the Sclera, according to COMS definition for the point of dose prescription[56] has been simulated in both phantoms. This tumor has been assumed to be on the lateral portion to the eyeball on the equator.

### C. Ophthalmic Plaque and Brachytherapy Source

In these investigations, a fully-loaded 16 mm COMS eye plaque, containing 13 $^{125}$I (model 6711, GE Healthcare/Oncura) brachytherapy sources have been modeled. These sources were sandwiched between a gold plaque with a density of 15.8 g/cm$^3$ and Silastic seed-career with a density of 1.12 g/cm$^3$. The geometric information and composition of the simulated plaque were obtained from some published references[57-62]. Also, the coordinates of the 13 seeds for this plaque are in accordance with the standard position for COMS-plaque[56]. Moreover, the detailed geometric and characteristic information of the model 6711 $^{125}$I sources have been obtained from the publicly accessible website of the Carleton Laboratory for Radiotherapy physics seeds database[63]. The photon spectra quoted in TG-43[54] has been used to sample the initial photon energies and probabilities for these brachytherapy sources. In these simulations, both water and human eyeball phantom with the fully loaded eye-plaque were placed in a 30×30×30 cm$^3$ water phantom, for the final dosimetric evaluations. The location of the eyeball in this phantom was selected such that it nearly represented the real patient anatomy (i.e., the cornea of the eye was toward one surface of the cubical phantom).

### D. Gold Nanoparticles (GNPs)

In our simulation, 50 nm GNPs were uniformly distributed within the tumor, in both eye phantoms filled with the water and the eye composition to create different concentrations (i.e. milligram of GNPs per gram of the target tissue) of 7, 10, 18 and 30 mg/g. From the simulated absorbed dose to different points of interests in the eyeball, with and without the presence of the GNPs, the values of the dose enhancement factors were calculated. Considering a treatment time of 100 hours, a procedure introduced by Thomson $et\ al^{44}$ was utilized to achieve a dose of 85 Gy at the tumor apex.

## III. Result
### A. Calculation of TG-43 Parameter (Single Source)

The accuracy of the seed model used in this study has been benchmarked via calculations of the TG-43 dosimetry parameters (Air Kerma Strength, dose rate constant and radial dose function) and comparison with the published results reported by Taylor $et\ al^{63}$ and M. J. Rivard $et\ al^{54}$.

Table 3 and Figure 2 shows excellent agreement (within ±5%) between the presently calculated radial dose function (RDF) of a single Oncoseed $^{125}$I (Model 6711) brachytherapy source and the reported data by Rivard $et\ al^{54}$ and Taylor $et\ al^{63}$.

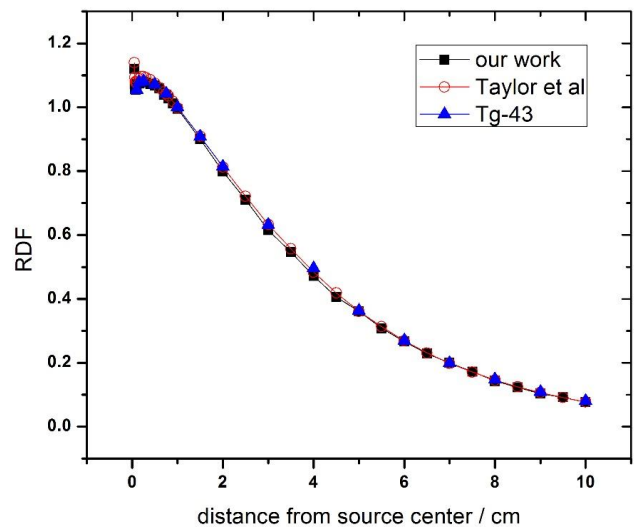

Figure 2: Radial dose function for 125I source. Voxel sizes are: 0.008 cm for distance between 0.05 < r ≤ 0.1, 0.01 cm for 0.1 < r ≤ 1 cm, 0.05 cm for 1 < r ≤ 5 and 0.1 cm for 5 < r ≤ 10. Symbol of Circle and Triangle are values calculated by Taylor et al$^{63}$ and J. Rivard et al$^{54}$.

Moreover, the dose rate at the reference point, ($r_0$ = 1 cm, $\theta_0$ = π/2) has been calculated, $\dot{D}$ ($r_0$, $\theta_0$), as the dose to water per photon history in a torus tally cells with 0.01cm minor cross sectional diameter. It should be noted that $R_0$ and $\theta_0$ are the transverse axis polar coordinates relative to the source center. The dose rate constant, Λ, has been obtained as the ratio of the calculated dose rate to the air kerma strength ($S_K$) with the unit of cGy.h$^{-1}$U$^{-1}$. The symbol U is the unit of air-kerma strength of the source and it is defined as 1U=1 cGy.cm$^2$.h$^{-1}$. The result gained from this study and those reported by TG-43[54] and Taylor et al[63] are listed in Table 4.

Figure 3 displays the isodose contours in the x, y plane at z=0 for a 16 mm COMS Plaque loaded with 13 $^{125}$I Oncoseeds (Model 6711). This plaque has been simulated at the center of the water phantom and the dose has been scored in (0.1 cm)$^3$ voxels. In this plot, the 100% line dose is located at the tumor apex. As the distance from this point increases, dose values decrease rapidly which is the main benefit of this treatment modality.

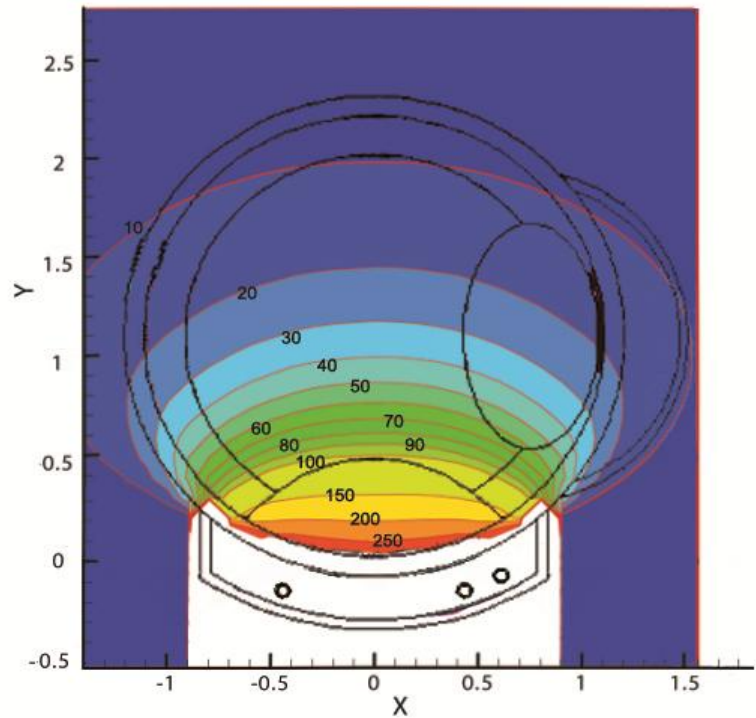

Figure 3: Z-plane, 16 mm plaque (in water) isodose lines from MCNP, 100% at tumor apex.

## B. Dosimetry calculations

Figure 4 (a) shows the depth dose curve for $^{125}$I sources in a fully-loaded 16 mm COMS eye plaque in both water and eye phantoms (without the presence of the tumor), relative to the dose value at 5 mm depth (where in this work tumor apex is assumed to be). These results show an agreement between the two sets of data for all the points except the first and last points that falls on the sclera adjacent to the plaque and opposite to the plaque, respectively. The dose value to the first and the end voxel in the eye phantom are approximately 26% higher than that of the voxels in water phantom. Based on the similarity between the elemental composition of the vitreous body and water, it is clear that the dose in the vitreous is approximately similar to that of the water. Moreover, the calculated depth dose in the water phantom has been compared with those calculated data by Thomson et al[44] (Figure 4 (b)) with an excellent agreement (within the uncertainties shown in Table 1) between them.

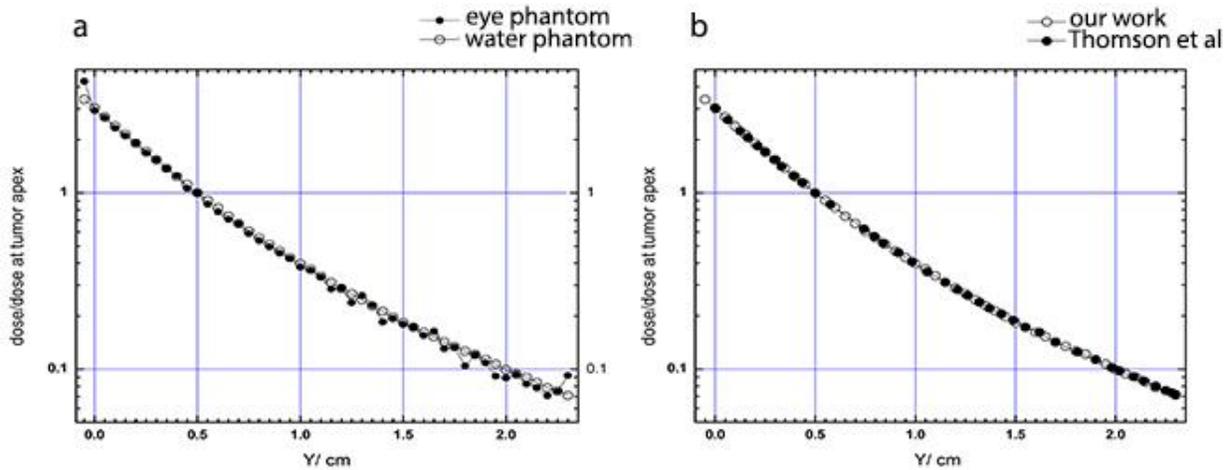

Figure 4: (a) The Plaque central axis depth-dose curves for I125 in the water and the eye phantom. (b) Comparison of the Plaque central axis depth-dose in the water phantom between this work and the work by Thomson et al[44]. The vertical axis represents the ratio of dose to the dose at the tumor apex.

Figure 4 shows the depth dose curves along the central axis direction of the plaque. To study the effect of the eyeball composition, the Monte Carlo simulation of the eye includes several (0.05 cm$^3$) voxels that have been selected at the different critical points in the eye as well as the tumor apex. With the assumptions and utilization of the approach of Thomson *et al*[44], the total dose at the points of interest have been calculated and they are shown in Table 5. The relative statistical uncertainty is lower than about 1% with the highest percent in the opposite side of the eye and the lowest percent in the sclera among the prescription points represented in this table. The calculated dose at points of interest for seeds in water and plaque in water has been compared with that reported by Thomson *et al*[44]. The air kerma strength per seed has been calculated for $^{125}$I model 6711 following the method of the air kerma strength per seed introduced by other investigators[43-44]. Figure 5 presents a comparison between 4 concentrations of GNPs in the calculation of dose enhancement factor (DEF) in both phantoms. The dose enhancement factor has been plotted in the figure as a function of radial distances from the center of the plaque while the plaque is placed next to the tumor in the simulated human eye and the water phantom.

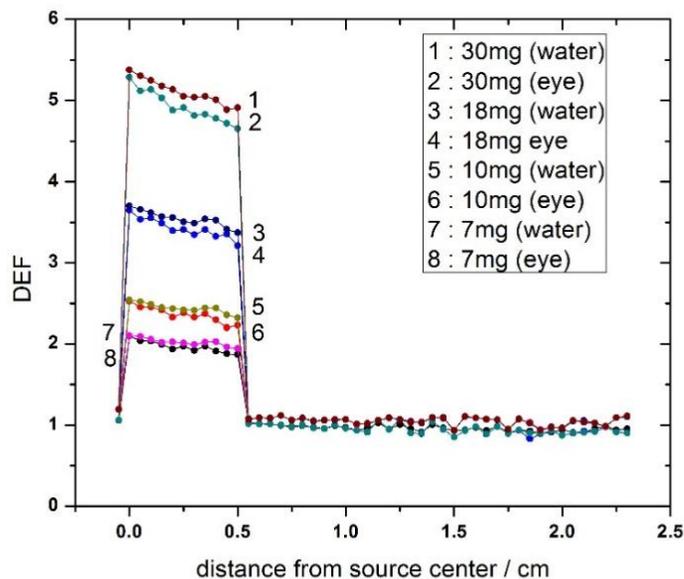

Figure 5: The calculated dose enhancement factors for 50nm GNPs within the tumor with concentrations of 7 mg/g, 10 mg/g, 18 mg/g, and 30mg/g in the water and the eye model phantoms. For Monte Carlo simulation the tally cells were placed along the central axis of the plaque, starting from the sclera near the plaque to the sclera opposite side of the plaque. DEF calculations have been done in the simulated human eye using the data shown in Table 2. Fully loaded 16 mm COMS eye plaque has been positioned next to the tumor on the equator temporal to the eyeball. A fully simulated human eye globe filled with eye material is referred to as (eye) and the water phantom is referred to as (water).

Figure 6 shows the ratio of the central axis depth dose to the dose at the tumor apex for 16mm COMS eye plaque in water phantom, eye globe and simulated human eye with nanoparticles-induced tumor. These results show that for the same dose to the apex of the tumor lower normal tissue doses (distances > 0.5 cm which is the tumor apex) are achieved for larger concentration of the nanoparticles.

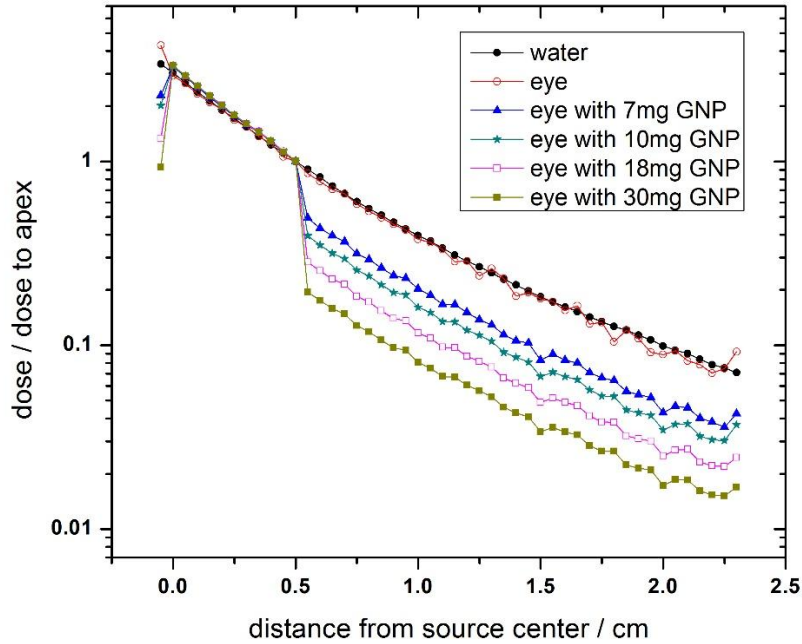

Figure 6: The ratio of the full loaded 16mm COMS eye plaque's depth doses to the dose at the tumor apex in the water phantom, simulated human eye, and simulated eye in which 4 concentrations of GNPs are defined inside the tumor.

## IV. Discussion

In radiation therapy, the required radiation dose for treatment is chosen based on the type of tumor and the rate of its progress. Determination of the required dose for the treatment of a specific tumor or the control of the tumor growth can be directly related to the treatment time for a given technique. For treatment of ocular melanoma, the prescribed dose is normally to the tumor apex. The height of the tumor apex may vary from patient to patient and thus the treatment time will vary, to deliver prescribed dose. These changes may affect the dose values to the normal tissues in the eye such as sclera, optical nerve, and lens. Table 5 show that defining the nanoparticles in the tumor area lead to the dose increase inside the tumor with no significant changes in the absorbed dose by other parts of the eye in both phantom types. The calculated doses to the tumor apex in the presence of GNPs inside the tumor with the concentrations of 7 mg/g, 10 mg/g, 18 mg/g, and 30 mg/g, was found to be increased by a factor of about 1.9, 2.2, 3.2 and 4.6, respectively, in the eye model and about 1.9, 2.3, 3.3 and 4.8 orders of magnitude, respectively, in the water phantom.

Despite the similarity of the dose enhancement to the tumor apex for the two phantom types, the results shown in Table 5 show that the composition of the eye materials influences the calculated dose in the choroidal melanoma. This effect is visible in the absence of GNPs, indicating that the dose to the points of interest in the eye phantom differs from that of water phantom. The dose to the center of lens and optic disk in the eye is lower than the dose calculated in these points in water. On the contrary, the dose to the sclera, apex and the opposite side in the eye are more than that of the water phantom. The dose increase in the sclera is about 28% and the dose reduction in the lens is about 11.6%.

The differences between the results of dosimetry in the eye model and water phantom are more prominent when the GNPs are defined in the tumor. For example, in the presence of 18mg/g GNPs there is an increase of 20% in the eye model and 4% in the water phantom in the dose absorbed to sclera relative to that of the eye and water phantom with no nanoparticles present. These differences are mainly due to the impact of the elemental compositions of the two phantom types in the photoelectric interaction of the low energy photons from $^{125}$I brachytherapy sources. Furthermore, with the presence of the plaque relative to the absence of the plaque, our results indicate that the dose to the point located opposite to the plaque, has an increase of approximately 10% in the water phantom while decreasing by about 7% in the eye phantom. This effect also can be attributed to the photoelectric interactions for the low energy photon and chemical elements of the two phantom types.

In table 5, a comparison of the effect of the plaque for multiple seed simulation has been shown too. The data shows that the presence of the plaque around the seeds causes a decrease in the dose to all points of interest. For instance, the dose to the sclera and tumor apex in the water phantom, in the presence of the plaque is about 14% less than that of the absence of the plaque, but with the same arrangement. This amount is increased as the distance from the plaque is increased. The dose reduction is a consequence of the elemental composition of the plaque backing gold alloy and the Silastic seed-career. The high atomic number of the gold causes enhanced photoelectric absorption and results in a dose reduction. Table 5 shows that, in the presence of the plaque the dose to sclera is approximately 28% larger in eye phantom than that of water phantom. The differences of the aforementioned increase of the dose values are due to the differences of the chemical compositions of the phantom material and presence of the plaque.

Regarding the results of Table 5 and considering the relative statistical uncertainty lower than about 1% in all calculations, the discrepancies between water and eye phantom are considerable in all parts of the eye. However, considering systematic uncertainties in eye plaque therapy, the discrepancies in some parts such as center of eye and Macula may be in the range of uncertainties, and therefore they may not be reliable.

As can be seen from the Figure 5, the tumor dose enhancement is greater for higher concentration of GNPs in both water and the eye phantoms. However, the notable result is that a given concentrations of GNPs, the DEF value in the water phantom is larger than that in the eye phantom. Moreover, as shown in this figure, the difference between the eye and the water phantom increases with increasing nanoparticle concentration. Since the photoelectric cross-section depends on the atomic number of the material and photon beam energy, the use of the $^{125}$I as a low energy photon source and GNP as a high Z material will increase the probability of the photoelectric interaction inside the tumor. Therefore, as shown in Figure 6, keeping a fixed prescribed dose delivery to the apex of the tumor will lead to a lower absorbed dose to the normal tissue of the eye. This reduction is increased by increasing the concentration of the GNPs inside the tumor.

## V. Conclusion

In this study, a Monte Carlo simulation model of human eye was built, considering its composition closer to reality, and a water phantom to investigate the effects of the GNPs on radiation dose enhancement in ophthalmic brachytherapy dosimetry. The results show that a significant tumor dose enhancement could be achieved, using GNPs inside the tumor during the irradiation by low energy source. With a certain diameter of GNPs, the results of the dose calculation show a higher dose enhancement for the greater concentration of GNPs.

The presence of GNPs inside the tumor made no significant changes in the radiation absorbing sensitivity of other normal tissue of the eye. Therefore, for the same delivery of the dose to the tumor apex, one may be able to deliver a smaller dose to the normal tissues in the eye (Figure 6). Furthermore, comparing the dosimetry calculations in the presence of GNPs between the water phantom and the eye model shows the importance of a more accurate definition of the eye material in ophthalmic brachytherapy.

The results of the Monte Carlo study in this investigation show that the presence of GNPs inside the tumor could play an important role in dose enhancement. However, the availability of an experimental study (in-vitro or in-vivo) in melanoma could result in a better understanding of the effect of GNP in the melanoma dosimetry. A full experimental investigation of the effects of the GNPs inside the choroidal melanoma on brachytherapy dosimetry could answer many of the questions in this project.

**Table 1**: Monte Carlo simulation uncertainties

| Component | r = 1 cm | r = 5 cm |
| --- | --- | --- |
| Statistics[a] | 0.5% | 1.0% |
| Photoionization | 1.5% | 4.5% |
| Cross section    (2.3%) | | |
| Seed geometry | 2.0% | 2.0% |
| Source energy spectrum [a] | 0.1% | 0.3% |
| Quadrature sum | 2.5% | 5.0% |

[a] On the transverse plane of a single source

Table 2: Elemental compositions and densities of different parts of the human eye-ball, water, and dry air used in this simulation.[44,47-52]

| Material | Material Elemental Composition (% by Mass) | | | | | | | | | | | | Density (g/cm$^3$) |
| --- | --- | --- | --- | --- | --- | --- | --- | --- | --- | --- | --- | --- | --- |
| | H | O | C | N | Na | Mg | P | S | Ar | Cl | K | Ca | |
| Water | 11.11901 | 88.8099 | --- | --- | --- | --- | --- | --- | --- | --- | --- | --- | 0.9980 |
| Dry air | --- | 23.17812 | 0.012425 | 75.52673 | --- | --- | --- | --- | 1.282725 | --- | --- | --- | 0.0012 |
| Lens | 9.60 | 64.60 | 19.50 | 5.70 | 0.10 | --- | 0.10 | 0.30 | --- | 0.10 | --- | --- | 1.07 |
| Sclera | 9.60 | 74.40 | 9.90 | 2.20 | --- | 0.50 | 2.20 | 0.90 | --- | 0.30 | --- | --- | 1.09 |
| Vitreous | 11.0867 | 88.052 | 0.068 | --- | 0.2647 | 0.0025 | 0.002 | --- | --- | 0.502 | 0.0171 | 0.005 | 1.00 |
| Aqoues humor | 11.085 | 88.06 | 0.056 | 0.00001 | 0.348 | 0.0025 | 0.0018 | 0.00001 | --- | 0.432 | 0.00798 | 0.0067 | 1.01 |
| Optic nerve | 10.70 | 76.70 | 9.50 | 1.80 | 0.20 | --- | 0.30 | 0.20 | --- | 0.30 | 0.30 | --- | 1.039 |
| Skull bone | 5.00 | 43.50 | 21.20 | 4.00 | 0.10 | 0.20 | 8.10 | 0.30 | --- | --- | --- | 17.60 | 1.61 |
| Tumor | 10.8 | 83.2 | 4.10 | 1.10 | --- | 0.30 | --- | 0.10 | --- | 0.40 | --- | --- | 1.03 |

**Table 3**: A comparison between the simulated radial dose function, g(r), of the oncoseed 6711 source in this project with the published data by other investigators.

| Distance from source (cm) | Radial Dose Function, g(r) | | |
|---|---|---|---|
| | This work | Taylor et al[63] | TG-43[54] |
| 0.05 | 1.115 | 1.139 | ---- |
| 0.06 | 1.050 | 1.094 | ---- |
| 0.07 | 1.068 | 1.08 | ---- |
| 0.08 | 1.069 | 1.077 | ---- |
| 0.09 | 1.071 | 1.078 | ---- |
| 0.10 | 1.050 | 1.08 | 1.055 |
| 0.15 | 1.064 | 1.089 | 1.078 |
| 0.20 | 1.066 | 1.096 | ---- |
| 0.25 | 1.062 | 1.096 | 1.082 |
| 0.30 | 1.059 | 1.093 | ---- |
| 0.40 | 1.050 | 1.086 | ---- |
| 0.50 | 1.075 | 1.075 | 1.071 |
| 0.60 | 1.060 | 1.062 | ---- |
| 0.70 | 1.040 | 1.048 | ---- |
| 0.75 | 1.040 | 1.042 | 1.042 |
| 0.80 | 1.030 | 1.035 | ---- |
| 0.90 | 1.012 | 1.018 | ---- |
| 1.0 | 0.995 | 0.998 | 1 |
| 1.5 | 0.900 | 0.909 | 0.908 |
| 2.0 | 0.800 | 0.813 | 0.814 |
| 2.5 | 0.710 | 0.721 | ---- |
| 3.0 | 0.615 | 0.633 | 0.632 |
| 3.5 | 0.550 | 0.557 | ---- |
| 4.0 | 0.472 | 0.484 | 0.496 |
| 4.5 | 0.406 | 0.419 | ---- |
| 5.0 | 0.361 | 0.361 | 0.364 |
| 5.5 | 0.310 | 0.313 | ---- |
| 6.0 | 0.267 | 0.269 | 0.270 |
| 6.5 | 0.230 | 0.231 | ---- |
| 7.0 | 0.199 | 0.198 | 0.199 |
| 7.5 | 0.172 | 0.171 | ---- |
| 8.0 | 0.143 | 0.146 | 0.148 |
| 8.5 | 0.123 | 0.125 | ---- |
| 9.0 | 0.104 | 0.107 | 0.109 |
| 9.5 | 0.092 | 0.0914 | ---- |
| 10.0 | 0.0762 | 0.0775 | 0.0803 |

**Table 4**: A comparison of the dose rate constant, $\Lambda$ (cGyU$^{-1}$h$^{-1}$), of $^{125}$I brachytherapy source in water, simulated in this project with the published data.

|  | R. Taylor, D. Rogers[63] | M. J. Rivard *et al*[54] | This work |
|---|---|---|---|
| Dose rate constant | 0.942 | 0.965 | 0.923±0.011 |

**Table 5**: A comparison of the integrated dose (Gy) at different points of interests in an eye-ball, implanted with a 16 mm eye plaque located next to the tumor on the equator temporal to the eyeball. Columns labeled "P.W" and "S.W" compares the dose for 13 $^{125}$I seeds in water phantom with and without the presence of the plaque, respectively. "Eye" refers to simulated human eye using the data of table 2 and "water" refers to the simulated human eye in which all parts of the eye were assumed to be made of water. "Plaque in eye" refers to the 13 $^{125}$I seeds with plaque present in human eye phantom. "7 mg/g, 10 mg/g, 18 mg/g, and 30 mg/g" refer to the concentration of GNPs inside the tumor in both phantom types. The relative statistical uncertainty is lower than 1%.

| location | Thomson et al[44] (S.W) | This work (S.W) | Thomson et al[44] (P.W) | This work (P.W) | Plque in eye | 7mg/g (water) | 7mg/g (eye) | 10mg/g (water) | 10mg/g (eye) | 18mg/g (water) | 18mg/g (eye) | 30mg/g (water) | 30mg/g (eye) |
|---|---|---|---|---|---|---|---|---|---|---|---|---|---|
| Sclera | 262.1 | 261.53 | 222.7 | 223.54 | 285.93 | 267.73 | 298.92 | 267.73 | 298.92 | 270.33 | 298.92 | 270.33 | 298.92 |
| Apex | 85.00 | 85.00 | 74.40 | 73.66 | 76.50 | 141.38 | 142.8 | 168.58 | 170.29 | 244.80 | 245.08 | 355.0 | 354.16 |
| Center of eye | 27.92 | 27.63 | 23.68 | 22.66 | 22.49 | 23.80 | 23.74 | 23.82 | 23.77 | 23.82 | 23.77 | 23.80 | 23.80 |
| Opposite side | 6.83 | 6.80 | 5.453 | 5.16 | 6.53 | 5.72 | 6.09 | 5.72 | 6.01 | 5.72 | 6.01 | 5.75 | 6.01 |
| Optic disk | 11.35 | 10.39 | 8.947 | 8.11 | 7.80 | 8.26 | 8.11 | 8.24 | 8.08 | 8.24 | 8.08 | 8.26 | 8.05 |
| Lens | 21.75 | 21.45 | 17.53 | 17.30 | 15.30 | 18.98 | 15.49 | 19.09 | 15.32 | 19.06 | 15.32 | 19.01 | 15.28 |
| Macula | 16.45 | 15.75 | 12.80 | 11.46 | 11.62 | 12.55 | 12.40 | 12.53 | 12.40 | 12.53 | 12.40 | 12.63 | 12.40 |